\documentstyle[amssymb,preprint,aps]{revtex}

\begin{document}
\title{Extraordinary Temperature Dependence of \\
the Resonant Andreev Reflection}
\author{{Yu Zhu, Qing-feng Sun and }Tsung-han Lin$^{*}$}
\address{{\it State Key Laboratory for Mesoscopic Physics and }\\
{\it Department of Physics, Peking University,}{\small \ }{\it Beijing}\\
100871, China}
\date{}
\maketitle

\begin{abstract}
An extraordinary temperature dependence of the resonant Andreev reflection
via discrete energy level in a normal-metal / quantum-dot / superconductor
(N-QD-S) system is predicted theoretically by using Green function
technique. The width of zero bias conductance peak in N-QD-S is about $\sqrt{%
\Gamma _L^2+\Gamma _R^2}$ and does not exhibit thermal broadening, where $%
\Gamma _L$ and $\Gamma _R$ are the coupling strength between QD and leads.
Considering the intra-dot Coulomb interaction, the Coulomb blockade
oscillations conducted by Andreev reflection differs dramatically from that
in N-QD-N. Instead of thermal broadening, finite temperature induces more
resonant peaks around the oscillation peaks of zero temperature. This effect
can be applied to determine the coupling strength and QD level spacing in
N-QD-S.
\end{abstract}



\baselineskip 20pt 

\newpage

Mesoscopic hybrid normal-metal / superconductor (N/S) systems have been
investigated intensively in the last decade. Aside from their potential
applications, these systems provide us an opportunity to combine with the
two different quantum coherent behaviors, the coherence of quasi-particles
in mesoscopic system and the coherence Cooper pairs in superconductors. The
key mechanism to connect the above two together is the Andreev reflection
(AR) process at the N/S interface \cite{Andreev,btk}. Many new phenomena and
effects involving AR\ have been addressed in recent years (for a review see 
\cite{review} and references therein). Among them, normal-metal /
quantum-dot / superconductor (N-QD-S) is one of the interesting systems, in
which resonant AR occurs via discrete energy levels of QD \cite
{rar1,rar2,ar1}. Considering the strong intra-dot interaction, the problem
of Kondo resonance in N-QD-S were also studied by several authors \cite
{kondo1,kondo2,kondo3,kondo4,kondo5}. In this paper, we shall predict the
unique temperature dependence of the resonant AR process and investigate the
interplay of Coulomb blockade effect and the Andreev tunneling in the N-QD-S
systems.

Temperature usually gives us the impression of averaging and smearing. At
finite temperature, the sharp resonant conductance peak usually has a
thermal broadening of the order of $k_BT$, but we find this is not true for
the case of resonant AR conductance. Let us begin by considering a N-QD-S
system modeled by the following Hamiltonian,

\begin{eqnarray}
H &=&H_L+H_R+H_{dot}+H_T\;\;, \\
H_L &=&\sum_{k\sigma }\epsilon _ka_{k\sigma }^{\dagger }a_{k\sigma }\;\;, 
\nonumber \\
H_R &=&\sum_{p\sigma }\epsilon _pb_{p\sigma }^{\dagger }b_{p\sigma
}+\sum_p\left( \Delta b_{p\uparrow }^{\dagger }b_{-p\downarrow }^{\dagger
}+h.c\right) \;\;,  \nonumber \\
H_{dot} &=&\sum_\sigma E_0c_\sigma ^{\dagger }c_\sigma \;\;,  \nonumber \\
H_T &=&\sum_{k\sigma }\left( t_La_{k\sigma }^{\dagger }c_\sigma +h.c\right)
+\sum_{p\sigma }\left( t_Rb_{p\sigma }^{\dagger }c_\sigma +h.c\right) \;\;, 
\nonumber
\end{eqnarray}
where $H_L$ and $H_R$ describe the left N lead and the right S lead, $%
H_{dot} $ describes QD with one spin-degenerate energy level, and $H_T$ is
the tunneling between them.

In the regime of $k_BT\ll \Delta $, the zero bias conductance is dominated
by the AR\ process, and can be derived by using Green function technique
(see \cite{ar1}) as, 
\begin{equation}
G_{NDS}=\frac{4e^2}h\int d\omega \left[ -\frac{\partial f(\omega )}{\partial
\omega }\right] \Gamma _L^2\left| G_{12}^r(\omega )\right| ^2\;\;,
\end{equation}
where $f(\omega )$ is the Fermi distribution function, $\Gamma _L$ is the
coupling strength (defined later), $G_{12}^r$ is the $12$ element of $%
2\times 2$ Nambu matrix ${\bf G}^r$, depicting the conversion from an
electron into a hole. ${\bf G}^r$ is defined by 
\begin{equation}
{\bf G}^r(\omega )=\int dt\;e^{-\text{i}\omega t}(-\text{i})\theta (t)\left( 
\begin{array}{ll}
<\{c_{\uparrow }(t),c_{\uparrow }^{\dagger }(0)\}> & <\{c_{\uparrow
}(t),c_{\downarrow }(0)\}> \\ 
<\{c_{\downarrow }^{\dagger }(t),c_{\uparrow }^{\dagger }(0)\}> & 
<\{c_{\downarrow }^{\dagger }(t),c_{\downarrow }(0)\}>
\end{array}
\right) \;\;,
\end{equation}
and can be solved from the Dyson equation, 
\begin{equation}
{\bf G}^r{\bf =g}^r{\bf +g}^r{\bf \Sigma }^r{\bf G}^r\;\;,
\end{equation}
where 
\begin{equation}
{\bf g}^r=\left( 
\begin{array}{cc}
\frac 1{\omega -E_0+\text{i}0^{+}} & 0 \\ 
0 & \frac 1{\omega +E_0+\text{i}0^{+}}
\end{array}
\right) \;\;,
\end{equation}

\begin{equation}
{\bf \Sigma }^r=-\frac{\text{i}}2\Gamma _L\left( 
\begin{array}{cc}
1 & 0 \\ 
0 & 1
\end{array}
\right) -\frac{\text{i}}2\Gamma _R\rho _R\left( 
\begin{array}{cc}
1 & -\frac \Delta \omega \\ 
-\frac \Delta \omega & 1
\end{array}
\right) \;,
\end{equation}
in which

\begin{equation}
\rho _R\equiv \left\{ 
\begin{array}{cc}
\frac{|\omega |}{\sqrt{\omega ^2-\Delta ^2}} & \;\;\;\;|\omega |>\Delta \\ 
\frac \omega {\text{i}\sqrt{\Delta ^2-\omega ^2}} & \;\;\;\;|\omega |<\Delta
\end{array}
\right. \;\;,
\end{equation}
and $\Gamma _{L/R}\equiv 2\pi N_{L/R}\left| t_{L/R}\right| ^2$, with $%
N_{L/R} $ being the density of states in the left/right lead in normal
state. For comparison we also write down the zero bias conductance for
N-QD-N \cite{wbl}, 
\begin{equation}
G_{NDN}=\frac{2e^2}h\Gamma _L\Gamma _R\int d\omega \left[ -\frac{\partial
f(\omega )}{\partial \omega }\right] \frac 1{(\omega -E_0)^2+\left( \frac{%
\Gamma _L+\Gamma _R}2\right) ^2}\;\;,
\end{equation}
in which single particle tunneling process dominates. In deriving the above
formulas, we assume that $\Gamma _{L/R}$ is independent of $\omega $, which
is reasonable in the calculation of zero bias conductance \cite{wbl}. In the
two limits of $k_BT\ll \Gamma _L,\Gamma _R$ and $k_BT\gg \Gamma _L,\Gamma _R$%
, the integrands of Eq.(2) and Eq.(8) can be evaluated, and the results are
summarized in the following table 
\begin{equation}
\begin{tabular}{|l|l|l|l|l|}
\hline
system & condition & conductance & half peak width & peak shape \\ \hline
N-QD-S & $k_BT\ll \Gamma _L,\Gamma _R$ & $4\frac{e^2}h\frac{\Gamma
_L^2\Gamma _R^2}{4\left( E_0^2+\frac{\Gamma _L^2+\Gamma _R^2}4\right) ^2}$ & 
$0.64\sqrt{\Gamma _L^2+\Gamma _R^2}$ & squared Lorentzian \\ \hline
N-QD-S & $k_BT\gg \Gamma _L,\Gamma _R$ & $4\frac{e^2}h\frac \pi {4k_BT}\frac{%
\Gamma _L\Gamma _R^2}{4\left( E_0^2+\frac{\Gamma _L^2+\Gamma _R^2}4\right) }$
& $\sqrt{\Gamma _L^2+\Gamma _R^2}$ & Lorentzian \\ \hline
N-QD-N & $k_BT\ll \Gamma _L,\Gamma _R$ & $2\frac{e^2}h\frac{\Gamma _L\Gamma
_R}{E_0^2+\left( \frac{\Gamma _L+\Gamma _R}2\right) ^2}$ & $\Gamma _L+\Gamma
_R$ & Lorentzian \\ \hline
N-QD-N & $k_BT\gg \Gamma _L,\Gamma _R$ & $2\frac{e^2}h\frac \pi {k_BT}\frac{%
e^{\beta E_0}}{\left( e^{\beta E_0}+1\right) ^2}\frac{2\Gamma _L\Gamma _R}{%
\Gamma _L+\Gamma _R}$ & $3.53k_BT$ & derivative of $f(\omega )$ \\ \hline
\end{tabular}
\end{equation}

Fig.1 shows the curves of $G_{NDS}$ vs $E_0$ and $G_{NDN}$ vs $E_0$ at
different temperatures for the symmetric coupling case $\Gamma _L=\Gamma
_R\equiv \Gamma $. One can see the following features: (1) For $T\rightarrow
0$, $G_{NDS}$ has its maximum $4e^2/h$ at $E_0=0$ while $G_{NDN}$ has $%
2e^2/h $, and $G_{NDS}$ is much steeper than $G_{NDN}$ which is of the
Lorentzian shape. (2) With the increase of temperature, both $G_{NDS}$ and $%
G_{NDN}$ are suppressed, but surprisingly, the conductance peak of $G_{NDS}$
does not exhibit thermal broadening while $G_{NDN}$ obviously does. Table
(9) tells us that the peak width of $G_{NDS}$ is always of the order $\sqrt{%
\Gamma _L^2+\Gamma _R^2}$ for either $k_BT\ll \Gamma _L,\Gamma _R$ or $%
k_BT\gg \Gamma _L,\Gamma _R$. Feature (1) has been obtained in \cite{rar1},
while feature (2) is addressed for the first time in this work, to our
knowledge.

Taking into account of that AR is a special two-particle process, feature
(2) has a simple interpretation shown in Fig.2. For N-QD-N (Fig.2b), the
conductance is dominated by the process of single-particle tunneling. At
finite temperature, the Fermi surface of leads spreads out around the
chemical potential in the range of $k_BT$, allowing incident electron has
energy $E$ in this range. The resonant tunneling occurs when $E_0=E$,
resulting in the broadening of conductance peak of the order $k_BT$. For
N-QD-S (Fig.2a), on the contrast, the single-particle tunneling is forbidden
due to superconducting gap, and AR\ process dominates. In the AR process, an
incident electron with energy $E$ (with respect to the superconductor
chemical potential) picks up another electron with energy $-E$ to form
Cooper pair and enter S, with a hole reflected in the Fermi sea of N. Since
both the incident electron and the picked electron pass QD\ through the
discrete energy level $E_0$, the resonant tunneling requires not only $E_0=E$
but also $E_0=-E$. Therefore, the resonant AR process occurs only when $%
E_0=0 $, i.e., $E_0$ lines up with the chemical potential of superconductor.
Because $E_0$ has a small broadening and shifting due to coupling with the
leads, the condition $E_0=0$ becomes to $\left| E_0\right| <O(\Gamma )$ with 
$O(\Gamma )$ being the order of the coupling strength. Thus, the resonant AR
causes a bottle-neck of the conductance peak width in N-QD-S, leading to a
broadening about $\Gamma $ instead of $k_BT$.

Next, we investigate a more realistic model, in which QD\ has multiple
single-particle energy levels and intra-dot Coulomb interaction, i.e., 
\begin{equation}
H_{dot}=\sum_{i\sigma }E_ic_{i\sigma }^{\dagger }c_{i\sigma }+\frac U2%
\sum_{i\alpha \neq j\beta }n_{i\alpha }n_{j\beta }\;\;,
\end{equation}
where $i,j$ are indices of the single-particle energy levels, $\sigma
,\alpha ,\beta $ are indices of the spin, and $U$ is the strength of Coulomb
interaction. For N-QD-N with intra-dot Coulomb interaction, it is well known
that the zero bias conductance vs the gate voltage exhibits Coulomb blockade
oscillations (CBO) \cite{cborev}. For N-QD-S in which one of N electrodes is
replaced by S, we also expect to see CBO except the conducting mechanism is
AR. Taking into account of the unusual temperature dependence of the
resonant AR process, CBO in N-QD-S may have significant change from that in
N-QD-N.

We constrain ourselves to discuss the weak coupling and low temperature
regime, i.e., $\Gamma \ll k_BT\ll \Delta $. If the coupling between QD and
leads is weak enough, the atomic limit solution ($\Gamma \rightarrow 0$) 
\cite{atomic} should be a good starting point. Because we study the zero
bias conductance, i.e., $V_L-V_R=0^{+}$, QD can be viewed as in equilibrium,
and the current driven by the small bias voltage servers as a probe to the
state of interacting QD. By generalizing the exact solution of the retarded
Green function of QD in the atomic limit (see the Appendix for details) , we
propose the following scheme for the calculation of zero bias conductance.
Suppose QD has $l$ spin-degenerate energy levels, $E_{1\uparrow
},E_{1\downarrow },E_{2\uparrow },E_{2\downarrow },\cdots E_{l\uparrow
},E_{l\downarrow },$ with $2^{2l}$ configurations of occupation represented
by $F=(N_{1\uparrow },N_{1\downarrow },N_{2\uparrow },N_{2\downarrow
},\cdots N_{l\uparrow },N_{l\downarrow })$ in which $N_{i\sigma }=0$ or $1$.
In the weak coupling regime, the retarded Green function of QD can be
derived by an approximated Dyson equation 
\begin{equation}
{\bf G}^r{\bf =\tilde{g}}^r{\bf +\tilde{g}}^r{\bf \Sigma }^r{\bf G}^r\;\;,
\end{equation}
in which ${\bf \tilde{g}}^r$ is the atomic limit solution for an isolated
interacting QD\ and ${\bf \Sigma }^r$ is the self-energy caused by the
tunneling between QD\ and leads. ${\bf \Sigma }^r$ is presented in Eq. (6),
and ${\bf \tilde{g}}^r$ can be obtained as 
\begin{eqnarray}
{\bf \tilde{g}}^r &\equiv &\sum_FP(F){\bf g}^r(F)\;\;, \\
P(F) &=&\frac 1Ze^{-\beta E_{dot}(F)}\;,\;Z=\sum_Fe^{-\beta E_{dot}(F)}\;\;,
\nonumber \\
{\bf g}^r(F) &=&\left( 
\begin{array}{cc}
\sum_i\frac 1{\omega -\tilde{E}_{i\uparrow }(F)+i0^{+}} & 0 \\ 
0 & \sum_i\frac 1{\omega +\tilde{E}_{i\downarrow }(F)+i0^{+}}
\end{array}
\right) \;\;,  \nonumber
\end{eqnarray}
in which $E_{dot}(F)\equiv \sum_{i\sigma }E_{i\sigma }N_{i\sigma }+\frac U2%
\sum_{i\alpha \neq j\beta }N_{i\alpha }N_{j\beta }$ is the total energy of
QD for the configuration $F$, and $\tilde{E}_{i\sigma }(F)\equiv E_{i\sigma
}+U\sum_{j\beta \neq i\sigma }N_{j\beta }$ is the renormalized sing-particle
levels of QD. And the total conductance through QD can be expressed in terms
of ${\bf G}^r$ derived from the above equations \cite{ar1}.

We plot $G_{NDS}$ vs $V_g$ and $G_{NDN}$ vs $V_g$ for different temperatures
in Fig.3a-3d and Fig.3e, where QD\ contains two spin-degenerate levels, $%
E_{1\uparrow }=E_{1\downarrow }=E_1=V_g$, $E_{2\uparrow }=E_{2\downarrow
}=E_2=V_g+\Delta E$, with level spacing $\Delta E=0.1$, interacting constant 
$U=1$. For $k_BT\ll \Delta E$, we see typical CBO pattern in Fig.3a, with
nearly equal spaced peaks at $V_g=0.0,\;1.0,\;2.1,\;3.1$. With the increase
of temperature, peak heights are suppressed , roughly proportional to $\frac %
1{k_BT}$ (Notice different scaling in Fig.3a-3d). The most striking feature
of CBO in N-QD-S is, when $k_BT\thicksim \Delta E$, temperature induces more
resonant peaks around the original CBO peaks, instead of thermal broadening.
One can see in Fig.3a-3d that the separation among peak groups is about $U$,
while the spacing of peaks within one peak group is $\frac{\Delta E}2$, and
each resonant peak has a width about $\Gamma $. Notice that all these
properties are independent on temperature.

Qualitatively, this unusual pattern can be understood as follows. Due to the
intra-dot Coulomb interaction, the original two resonances $E_1=E_{1\uparrow
}=E_{1\downarrow }$ and $E_2=E_{2\uparrow }=E_{2\downarrow }$ exhibits eight
sub-resonances, $E_1$,\ $E_2$,\ $E_1+U$, $E_2+U$, $E_1+2U$, $E_2+2U$, $%
E_1+3U $, $E_2+3U$ (see the appendix and Ref.\cite{cbothy} for detail). At
zero temperature, only four resonances $E_1$,$\;E_1+U$, $E_2+2U$, $E_2+3U$
are active, and the other four are Coulomb blockaded. But for higher
temperature of $k_BT\thicksim \Delta E$, the Coulomb blockade effect is
partially removed. Not only $E_1$ but also $E_2$ are active and contribute
to the total conductance. For N-QD-N, since resonant peak is broadened by
finite temperature, the contributions of $E_1$ and $E_2$ are
indistinguishable and combined to form the first CBO peak around $V_g=0$ in
Fig.3e. (This mechanism was first proposed by Y. Meir ${\sl et}$ ${\sl al}$.
to explain the anomalous temperature dependence of CBO peak heights in
N-QD-N \cite{cbothy}.) For N-QD-S, however, resonant AR peak has the width
of $\Gamma $ instead of $k_BT$. The first group of peaks around $V_g=0$ in
Fig.3c consists of three distinguishable contributions of AR: AR via $E_1$,
AR via $E_2$, and AR between $E_1$ and $E_2$ \cite{ar1}. Other groups of
peaks can be understood similarly. Furthermore, one find groups with tiny
peaks around $V_g=0.5,\;1.5,\;2.5$ in Fig.3d, which can be attributed to the
AR process between $E_1$ and $E_1+U$, $E_1$ and $E_2+U$, $E_2$ and $E_2+U$, $%
E_2$ and $E_1+U$, etc. In short, temperature dependence of CBO in N-QD-S
differs dramatically from that in N-QD-N.

Finally, we would like to make three remarks: (1) The extraordinary
temperature dependence of CBO in N-QD-S provides a new approach to determine
the coupling strength $\Gamma \thicksim \sqrt{\Gamma _L^2+\Gamma _R^2}$ and
QD level spacing $\Delta E$, which are impossible to do in N-QD-N because of
thermal broadening. (2) The suggested N-QD-S structure is accessible of the
up-date nano-technology, either by a metal nano-particle connected to
superconducting and normal metal electrodes \cite{nanopart1,nanopart2}, or
in a gate controlled N-2DEG-S structure \cite{2deg1,2deg2,2deg3}. The key
point to observe the predicted unusual temperature dependence of AR\ is to
perform resonant AR process via discrete energy levels. (3) Similar
extraordinary temperature dependence of zero bias conductance have been
observed experimentally either in normal-metal / semiconductor /
superconductor junctions \cite{exp1} or in normal-metal / high-T$_c$
superconductor contacts \cite{exp2}. Perhaps, the discrete electronic state
in our model corresponds to the quantum well state and surface bound state
in their situations.

In conclusion, we predict an extraordinary temperature dependence of the
resonant AR process in N-QD-S, in which the conductance peak does not
exhibit thermal broadening. We also investigate CBO in N-QD-S, and find that
finite temperature induces more resonant peaks around the original CBO peaks
of zero temperature, rather than gives them thermal broadening, which can be
applied to determine $\Gamma $ and $\Delta E$ in N-QD-S.

This project was supported by NSFC\ under Grant No. 10074001. T. H. Lin
would also like to thank the support from the Visiting Scholar Foundation of
State Key Laboratory for Mesoscopic Physics in Peking University.

\smallskip $^{*}$ To whom correspondence should be addressed.

\newpage

\section*{Appendix}

In this appendix, we present the atomic solution in equilibrium, which is
generalized to the weak coupling case and linear response regime in section
II.

In the atomic limit, QD\ is nearly isolated from the leads, and the
negligible coupling to leads determines the equilibrium / non-equilibrium
distribution in QD. Suppose QD\ has multiple discrete energy levels, indexed
by $i=1,2,\cdots L$, (here $i$ contains the spin index), and QD\ in the
atomic limit can be described by 
\begin{equation}
H_{dot}=\sum_{i=1}^LE_ic_i^{\dagger }c_i+U\sum_{i<j}n_in_j\;\;,  \eqnum{A1}
\end{equation}
in which the second term is the intra-dot Coulomb interaction, $U\equiv
e^2/2C$ is the charging energy of QD.

By using the equation of motion, the retarded Green function of QD\ can be
solved exactly. For example, for a three-level QD, $H_{dot}=%
\sum_{i=1}^3E_ic_i^{\dagger }c_i+U(n_1n_2+n_2n_3+n_3n_1)$, one obtains 
\begin{equation}
\langle \langle c_1|c_1^{\dagger }\rangle \rangle ^r=\frac{\langle
(1-n_2)(1-n_3)\rangle }{\omega -E_1}+\frac{\langle (1-n_2)n_3\rangle }{%
\omega -E_1-U}+\frac{\langle n_2(1-n_3)\rangle }{\omega -E_1-U}+\frac{%
\langle n_2n_3\rangle }{\omega -E_1-2U}\;\;.  \eqnum{A2}
\end{equation}
($\omega $ contains an infinitesimal imaginary part i0$^{+}$, for all the
retarded Green functions). The atomic limit solution have clear physics
meaning: due to intra-dot interaction, the single particle energy level $E_i$
exhibits several sub-resonances, $E_i,\;E_i+U,\;E_i+2U\cdots ,$ the weights
of these sub-resonances are determined by the occupation configuration of
other energy levels. For the three-level QD, $E_1$ has the weight of $%
\langle (1-n_2)(1-n_3)\rangle $, i.e., the probability of both $E_2$ and $%
E_3 $ are empty; $E_1+U$ has the weights of $\langle (1-n_2)n_3\rangle $ and 
$\langle n_2(1-n_3)\rangle $, i.e., the probability of one of $E_2$ and $E_3$
is occupied and the other empty; $E_1+2U$ has the weight of $\langle
n_2n_3\rangle $, i.e., the probability of both $E_2$ and $E_3$ are occupied.

The calculation of the correlation functions like $\langle n_1n_2\rangle $
should be done self-consistently when QD is in the non-equilibrium
distribution. However, for the equilibrium QD in the atomic limit, the
calculation is straight forward. In this case, the thermal equilibrium QD\
can be depicted by the density operator $\rho =\frac 1Ze^{-\beta H_{dot}}$,
and the observable $\langle O\rangle $ can be calculated by $Tr\langle \rho
O\rangle $. For the three-level QD, the results are listed as follows 
\begin{equation}
\begin{tabular}{|l|l|l|l|}
\hline
$\langle n_1n_2n_3\rangle $ & $\frac 1Ze^{-\beta (E_1+E_2+E_3+3U)}$ & $%
\langle n_1(1-n_2)(1-n_3)\rangle $ & $\frac 1Ze^{-\beta E_1}$ \\ \hline
$\langle (1-n_1)n_2n_3\rangle $ & $\frac 1Ze^{-\beta (E_2+E_3+U)}$ & $%
\langle (1-n_1)n_2(1-n_3)\rangle $ & $\frac 1Ze^{-\beta E_2}$ \\ \hline
$\langle n_1(1-n_2)n_3\rangle $ & $\frac 1Ze^{-\beta (E_1+E_3+U)}$ & $%
\langle (1-n_1)(1-n_2)n_3\rangle $ & $\frac 1Ze^{-\beta E_3}$ \\ \hline
$\langle n_1n_2(1-n_3)\rangle $ & $\frac 1Ze^{-\beta (E_1+E_2+U)}$ & $%
\langle (1-n_1)(1-n_2)(1-n_3)\rangle $ & $\frac 1Ze^{-\beta \cdot 0}$ \\ 
\hline
\end{tabular}
\eqnum{A3}
\end{equation}
And the weights in Eq.(A2) can be evaluated with the help of this table, for
example, $\langle n_2n_3\rangle =$ $\langle (1-n_1)n_2n_3\rangle +\langle
n_1n_2n_3\rangle $.

Generally, QD with $L$ energy levels may have $2^L$ occupation
configurations, represented by $F=(N_1,N_2\cdots N_L)$ with $N_i=0$ or $1$.
One has 
\begin{eqnarray}
P(F) &\equiv &\left\langle \prod_{i=1}^Lm_i\right\rangle =\frac 1Ze^{-\beta
E_{dot}(F)}\;\;  \eqnum{A4} \\
m_i &\equiv &\left\{ 
\begin{array}{cc}
n_i & \;\;for\;N_i=1 \\ 
(1-n_i) & \;\;for\;N_i=0
\end{array}
\right. \;\;,  \nonumber
\end{eqnarray}
in which $E_{dot}(F)\equiv \sum_iE_iN_i+U\sum_{i<j}N_iN_j$ is the total
energy of QD for the configuration $F$. Define $\tilde{g}^r\equiv
\sum_{i=1}^L\langle \langle c_i|c_i^{\dagger }\rangle \rangle ^r$, it is
easy to obtain 
\begin{eqnarray}
\tilde{g}^r &=&\sum_FP(F)g^r(F)  \eqnum{A5} \\
g^r(F) &=&\sum_{i=1}^L\frac 1{\omega -\tilde{E}_i(F)}\;\;,  \nonumber
\end{eqnarray}
in which $\tilde{E}_i(F)\equiv E_i+U\sum_{j\neq i}N_j$ is the renormalized
sing-particle level.

In short, in the atomic limit, the retarded Green function $\tilde{g}^r$ is
the average over different occupation configurations weighted by a thermal
factor, and each configuration behaves as a set of renormalized single
particle levels.


\newpage

\section*{Figure Captions}

\begin{itemize}
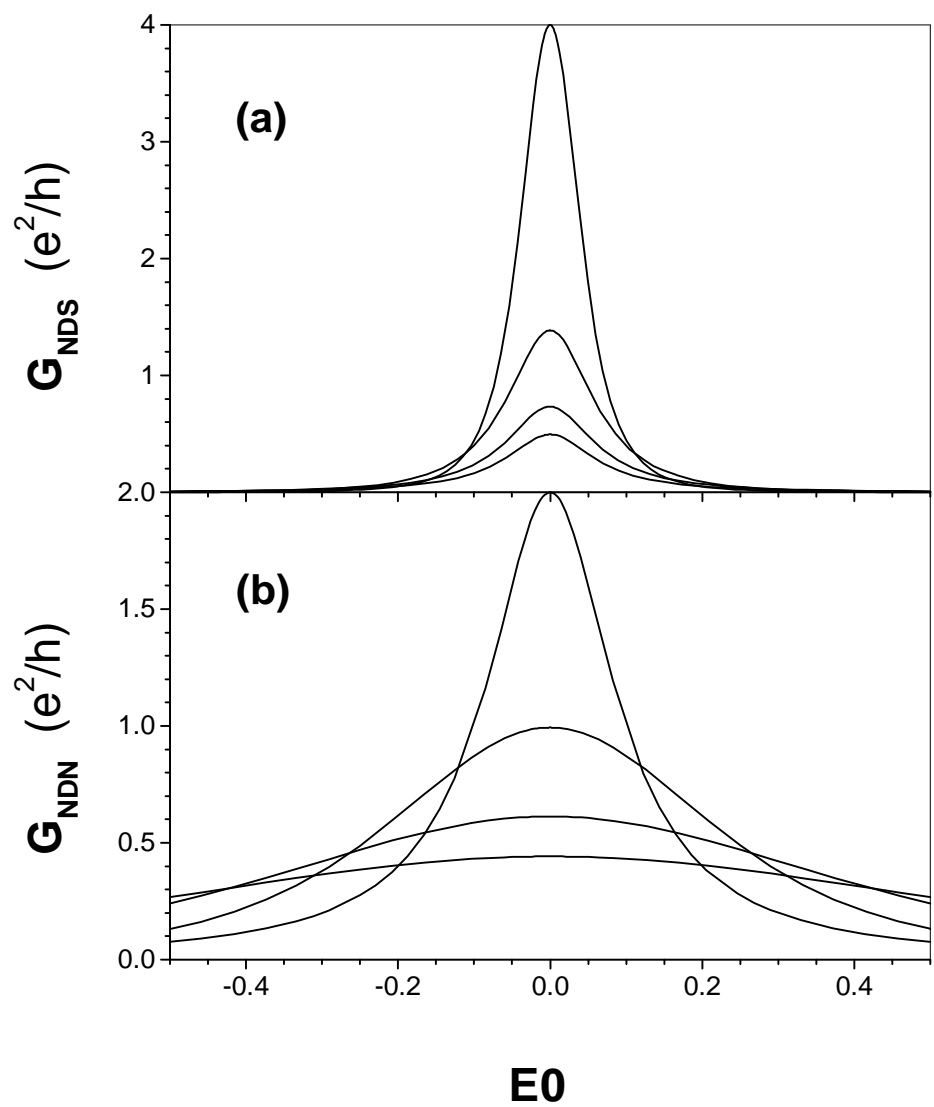
\item[{\bf Fig. 1}]  The zero bias conductance $G$ vs the resonant level $%
E_0 $ at different temperature for (a) N-QD-S and (b) N-QD-N. Parameters
are: $\Gamma _L=\Gamma _R=0.1$,$\;\Delta =1$ for N-QD-S, $\Delta =0$ for
N-QD-N, $k_BT=0.001,\;0.1,\;0.2,\;0.3$ corresponding to the decrease of peak
heights.

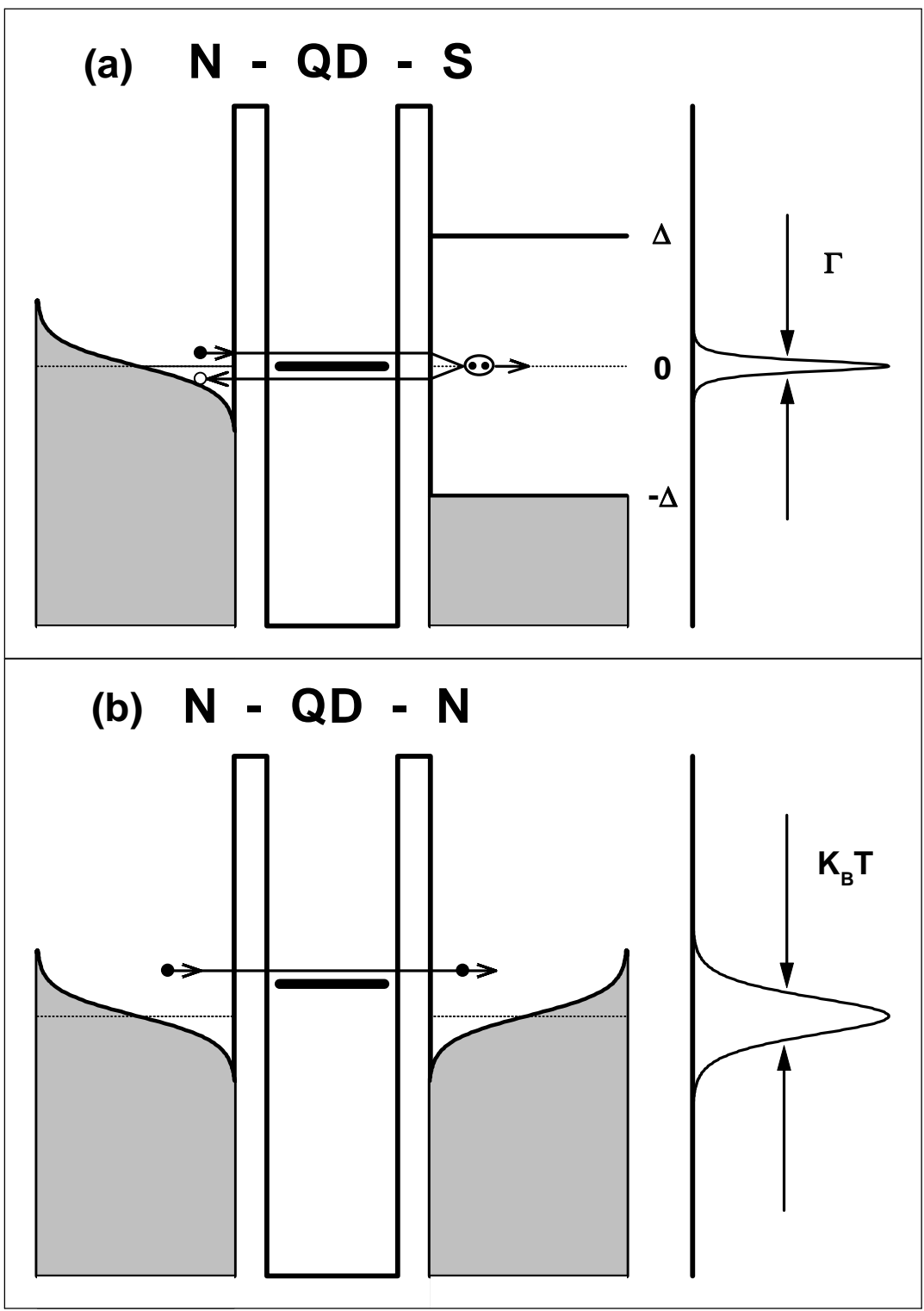
\item[{\bf Fig. 2}]  Schematic picture for understanding the unusual
temperature dependence of the conductance in Fig.1. For (a) N-QD-S, resonant
AR\ requires $E_0$ lines up with the chemical potential of the
superconducting lead. For (b) N-QD-N, single-particle tunneling occurs when
the resonant level $E_0$ within the range of $k_BT$.

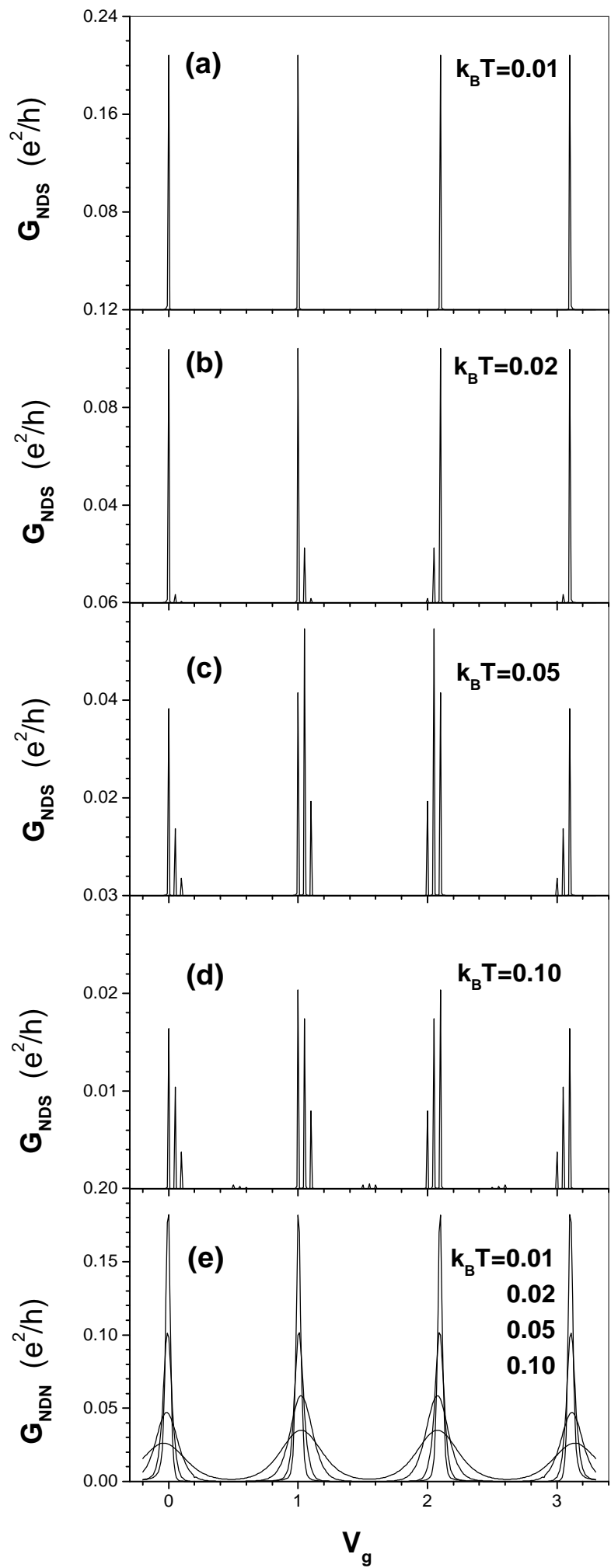
\item[{\bf Fig. 3}]  Coulomb blockade oscillations of $G$ vs $V_g$ at
different temperatures for N-QD-S in Fig.3a-3d and N-QD-N in Fig.3e. QD\ has
two spin degenerate levels, $E_{1\uparrow }=E_{1\downarrow }=E_1=V_g$ and $%
E_{2\uparrow }=E_{2\downarrow }=E_2=V_g+\Delta E$, with level spacing $%
\Delta E=0.1$, Coulomb interacting strength $U=1$. Other parameters are: $%
\Gamma \equiv \Gamma _1=\Gamma _2=0.002$, $\Delta =1$ for N-QD-S, $\Delta =0$
for N-QD-N, $k_BT=0.01,\;0.02,\;0.05,\;0.10$ marked in the plots.
\end{itemize}

\end{document}